\documentclass{elsart}

\usepackage{natbib}

   \usepackage{graphicx}
  \usepackage{amssymb}

\def\teq#1{$\, #1\,$}                         % text equation
\def\aa{{Astron. Astrophys.}}
\def\aap{{Astron. Astrophys.}}

\def\apj{ApJ}

\def\nat{Nature}
\def\app{Astroparticle Phys.}                   % DO NOT DELETE
\def\jetp{Sov. Phys. JETP}                      % DO NOT DELETE

\def\pasj{{Pub. Astron. Soc. Japan}}
\def\ssr{Space Sci. Rev.}                       % DO NOT DELETE
\def\reference{\par \noindent \hangafter=1 \hangindent=0.7 true cm}
             \font\sevenrm=cmr7

          \font\sixrm=cmr6

\def\tSNR{{t}_{\hbox{\sixrm SNR}}}
\def\dover#1#2{\hbox{${{\displaystyle#1 \vphantom{(} }\over{
   \displaystyle #2 \vphantom{(} }}$}}

\begin{document}
\newcommand{\vol}[2]{$\,$\rm #1\rm , #2.}                 
\def\gamsk{\gamma_1}
\def\erg{\varepsilon_\gamma}
\def\dover#1#2{\hbox{${{\displaystyle#1 \vphantom{(} }\over{
   \displaystyle #2 \vphantom{(} }}$}}
\def\thetascatt{\theta_{\hbox{\sevenrm scatt}}}
\def\thetaBone{\Theta_{\hbox{\sevenrm Bn1}}}
\def\thetaBtwo{\Theta_{\hbox{\sevenrm Bn2}}}
{\catcode`\@=11 
  \gdef\SchlangeUnter#1#2{\lower2pt\vbox{\baselineskip 0pt\lineskip0pt 
  \ialign{$\m@th#1\hfil##\hfil$\crcr#2\crcr\sim\crcr}}}} 
\def\gtrsim{\mathrel{\mathpalette\SchlangeUnter>}} 
\def\lesssim{\mathrel{\mathpalette\SchlangeUnter<}} 
\newcommand{\figureoutpdf}[5]{\centerline{}
   \centerline{\hspace{#3in} \includegraphics[width=#2truein]{#1}}
   \vspace{#4truein} \caption{#5} \centerline{} }
\begin{frontmatter}



\title{Multiwavelength Spectral Models\\ 
for SNR G347.3-0.5 from\\
Non-Linear Shock Acceleration}

\author{Matthew G. Baring$^a$, 
              Donald C. Ellison$^b$, and
              Patrick O. Slane$^c$}

\address{$^a$Department of Physics and Astronomy, MS-108,
   Rice University, P. O. Box 1892, Houston, TX 77251-1892, USA,
   {\tt Email: baring@rice.edu}\\
   $^b$Department of Physics, North Carolina State University,
   Box 8202, Raleigh, NC 27695, USA,
   {\tt Email: don\_ellison@ncsu.edu}\\
   $^c$Harvard-Smithsonian Center for Astrophysics, 
   60 Garden Street, Cambridge, MA 02138, USA,
   {\tt Email: slane@head.cfa.harvard.edu}}

\begin{abstract}
The remnant G347.3-0.5 exhibits strong shell emission in the radio and
X-ray bands, and has a purported detection in the TeV gamma-ray band by
the CANGAROO-II telescope.  The CANGAROO results were touted as evidence
for the production of cosmic ray ions, a claim that has proven
controversial due to constraining fluxes associated with a proximate
unidentified EGRET source 3EG J1714-3857.  HESS has now seen this source
in the TeV band. The complex environment of the remnant renders modeling
of its broadband spectrum sensitive to assumptions concerning the nature
and parameters of the circumremnant medium.  This paper explores a sampling
of reasonable possibilities for multiwavelength spectral predictions
from this source, using a non-linear model of diffusive particle
acceleration at the shocked shell.  The magnetic field strength, shell
size and degree of particle cross-field diffusion act as variables to
which the radio to X-ray to gamma-ray signal is sensitive.  The modeling
of the extant data constrains these variables, and the potential impact
of the recent HESS detection on such parameters is addressed.  Putative
pion decay signals in hard gamma-rays resulting from hadronic
interactions in dense molecular clouds are briefly discussed; the
requisite suppression of the GeV component needed to accommodate the 3EG
J1714-3857 EGRET data provides potential bounds on the diffusive
distance from the shell to the proximate clouds.
\end{abstract}


\end{frontmatter}
\setlength{\parindent}{.25in}

\section{Introduction}
\label{sec:Introduction}

Supernova remnants (SNRs) are believed to be the principal candidates for
generating galactic cosmic rays, relativistic ions and electrons with
energies from the GeV range up to \teq{10^{15}}eV and beyond.  A Holy
Grail of cosmic ray physics has been to secure a radiative detection
that would unambiguously indicate the presence of cosmic ray ions in
proximity to one or more supernova remnants, thereby establishing a
direct connection between source and diffuse population.  The search has
therefore focused on gamma-ray emission signaling the decay of neutral
pions, \teq{\pi^0\to\gamma\gamma}; these are presumably generated in
collisions between cosmic ray protons accelerated at the shocked outer
shells of remnants and ambient hydrogen.  This is a difficult task, due
largely to the relatively low fluxes expected, which approach the limits
of sensitivity in the 100 MeV band for past hard gamma-ray experiments
such as EGRET.

While this will be addressed by the upcoming GLAST mission, scheduled
for launch in 2007, the Imaging Atmospheric \v{C}erenkov Telescope
(IACT) technique has focused a large amount of time and energy over the
last decade in searching for pion decay signatures in the TeV band. 
Seminal steps along this path were the announcement of detections of the
young remnant SN1006 by CANGAROO (Tanimori et al. 1998), and even
younger Cas A by HEGRA (Aharonian et al. 2001) in the northern
hemisphere.  A second southern remnant, G347.3-0.5 (radio designation,
or RX J1713.7-3946 in X-rays), has a claimed detection by the CANGAROO
team (Enomoto et al. 2002) in its northwest rim, an asymmetry mirroring
the CANGAROO TeV observation of SN1006 (which contrasts its beautifully
symmetric radio and X-ray maps: e.g., see Koyama et al. 1995). The RX
J1713.7-3946 discovery paper spawned a mini-controversy over whether or
not the detection could be interpreted as evidence for hadrons in the
remnant. The model interpretation of this remnant forms the focus of
this work.

The CANGAROO team offered a hadronic interpretation for the signal they
claimed in Enomoto et al. (2002), using particle distributions that
would result from linear shock acceleration models. This assessment was
disputed in subsequent papers in the literature, in particular Reimer \&
Pohl (2002) and Butt et al. (2002).  These dissenting opinions argued
that in order to fit the spectrum published in Enomoto et al. (2002) in
the TeV range, the model pion decay flux at 100 MeV energies would be so
large that it would far exceed that of the EGRET unidentified source 3EG
J1714-3857 lying in the field of view of G347.3-0.5.   Whether or not
the EGRET source is associated with the remnant is immaterial: it can
serve as a putative upper bound, suggesting that an inverse Compton
model using cosmic microwave background (CMB) seed photons would be a
more viable explanation.

This paper offers additional perspectives on this source, specifically
exploring predictions of radiation models that incorporate the physics
of non-linear acceleration at remnant shocks.  Essentially it updates
the careful exploration of Ellison, Slane and Gaensler (2001) that
predated the full publication of the CANGAROO data on this source,
providing elements of a parameter survey.  An update on radio and X-ray
observations for this remnant has recently been provided by Lazendic et
al. (2004).  Yet crucial to multi-wavelength modeling is a recent
observational development: the new stereoscopic HESS telescope in
Namibia has now detected G347.3-0.5 with an image that nicely matches
the X-ray map, albeit with a slightly lower flux and a flatter spectrum
from the CANGAROO claim. The HESS results are presented in Aharonian et
al. (2004), and complicate the modeling picture considerably,
particularly since HESS has not detected SN 1006 (Aharonian et al.
2005), providing upper limits an order of magnitude or more below the
CANGAROO data.  The model results presented here for G347.3-0.5 will
leave the validity of the different TeV data sets to further appraisal
by the observational community.

\section{Multi-wavelength Models from Non-linear Shocks}
 \label{sec:non-linear}
 
Model predictions for supernova remnant emission often invoke linear
models of acceleration, where the accelerated particles have no
influence on the hydrodynamics of the expanding shock (e.g. Drury,
Aharonian \& V\"olk 1994; Gaisser, Protheroe \& Stanev 1998; Sturner et
al. 1997; Enomoto, et al. 2002; Reimer \& Pohl 2002).  Such linear
invocations generally suffice to predict X-ray and TeV fluxes to within
factors of \teq{3-10}.  Usually, acceleration is considered to take
place at the expanding forward shock, though increasingly the importance
of acceleration at the reverse shock is being realized (e.g.
Decourchelle, Ellison \& Ballet 2000; Ellison, Decourchelle \& Ballet
2005). Acceleration in strong forward shocks can be very efficient just
prior to, or in the early stages of, the Sedov phase, leading to a
sizeable fraction of the total energy budget being placed into cosmic
rays. Accordingly, energy flux conservation in the shock hydrodynamics
is profoundly influenced by the contribution from the highest energy
cosmic rays, modifying the shock compression, and hence inducing a
non-linear feedback on the acceleration process.

Such non-linear acceleration effects are well-documented in the cosmic
ray literature (e.g. see Drury 1983; Jones \& Ellison 1991 for reviews).
 Their key characteristic, a direct consequence of diffusive spatial
scales generally being an increasing function of particle momenta, is a
deviation from pure power-law behavior in the ion and electron
distributions, with an upward curvature indicating more efficient
acceleration of the most energetic particles.  Such modifications to
simpler, linear (test particle) models can influence radiative
predictions for comparison with data by factors of \teq{3-10} or more,
and therefore become imperative for the theory/observation interface. 
Detailed models of non-linear acceleration come in different varieties:
Monte Carlo simulations have been documented in Ellison \& Eichler
(1984), Jones \& Ellison (1991) and Ellison, Baring \& Jones (1996), and
analytic/numerical techniques involving diffusion-convection equations
have been developed by Kang \& Jones (1995), Berezhko, Yelshin \&
Ksenofontov (1996), Malkov (1997) and Blasi (2002).
Extensively-developed applications of these complementary approaches to
SNR emission yield similar radiative predictions for similar parameters,
so that there is a robustness in the theoretical modeling (e.g. see
Baring 2000, for a discussion).

In this paper, a rather compact analytic representation of the
non-linear problem developed for ions in Berezhko \& Ellison (1999), and
extended to treat electrons in Ellison, Berezhko \& Baring (2000), is
used.  This formulation models the non-linear feedback and uses
piecewise continuous broken power-laws to approximate the non-thermal
distributions.  The indices of the power-law segments are determined
self-consistently, and the resulting distributions are extremely facile
for folding into radiation models for remnants.  The reader is referred
to Ellison, Berezhko \& Baring (2000) for a comprehensive discussion
of the acceleration theory,
and Ellison, Slane \& Gaensler (2001) for a compact presentation
on its application to G347.3-0.5.

\subsection{Leptonic Spectral Models}
 \label{sec:leptonic}
 
In the light of the gamma-ray constraints imposed by the EGRET source in
the locale of G347.3-0.5 on the sky, it is prudent to focus on inverse
Compton models for TeV photon production, following Reimer \& Pohl
(2002), and many authors in applications to other remnants.  The radio
to X-ray continuum is modeled with synchrotron emission from a
shock-accelerated distribution of electrons, and the same distribution
upscatters CMB photons to gamma-ray energies.  In such models,
spectroscopically, ions go along for the ride, their major role being a
power drain for radiative efficiency.  Photon production rates are
standard calculations, assuming isotropy of electrons, and the reader is
referred to Baring et al. (1999), for example,  for details.

\begin{figure}[htb]
\figureoutpdf{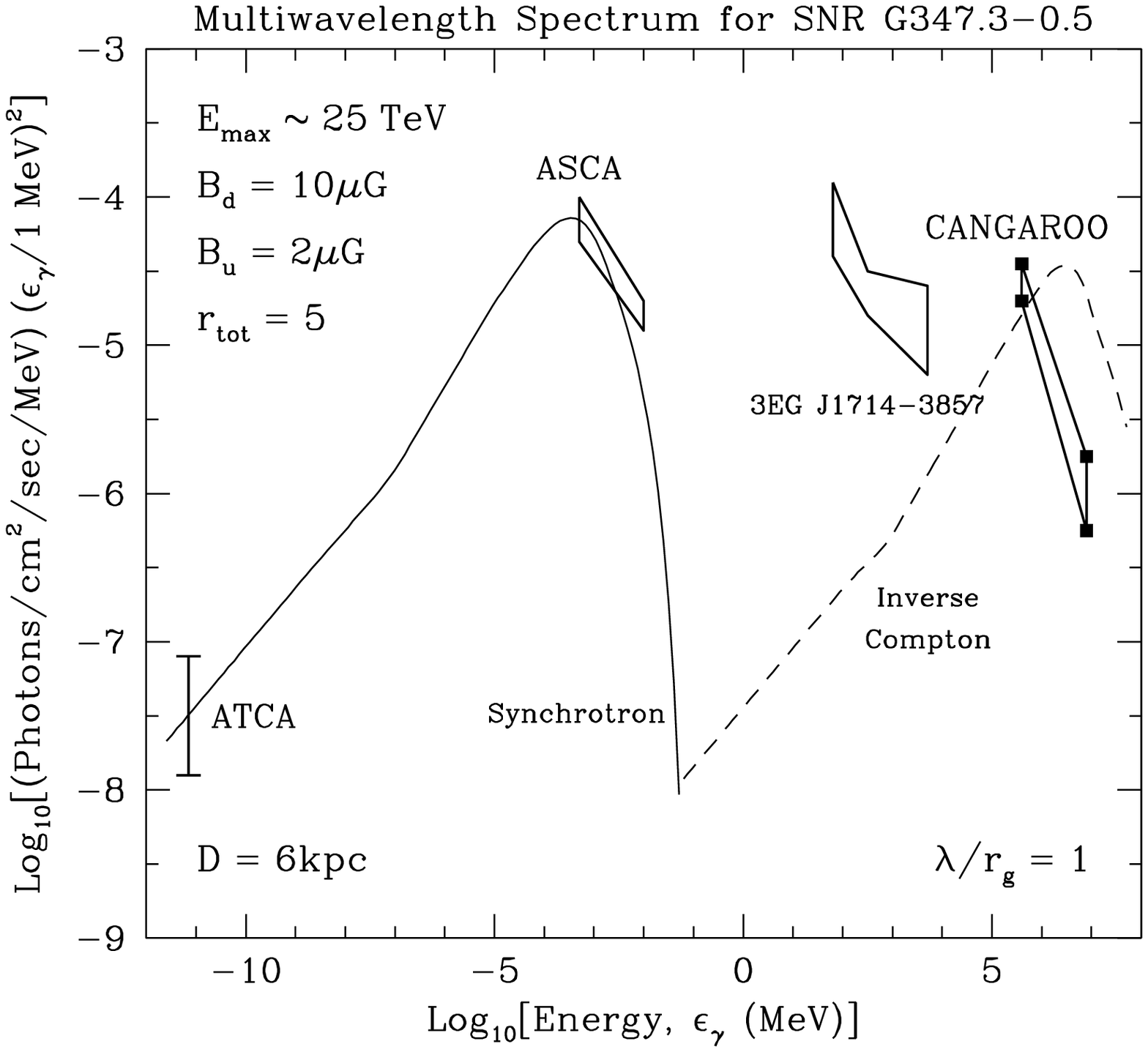}{5.5}{0.0}{-0.25}{
The multi-wavelength spectra for synchrotron (solid curve) and inverse 
Compton (dashed) components resulting from a non-linear shock acceleration
model of SNR G347.3-0.5 (RX J1713.7-3946).  Model/source parameters are
as labeled, and are discussed in the text.   The observational data
consists of 1.36 GHz radio observations by ATCA, a 2--10 keV ASCA X-ray
spectrum of the northwest rim (both detailed in Ellison, Slane \&
Gaensler 2001), a band at \teq{>300}GeV depicting the steep CANGAROO-II
spectrum announced in Enomoto et al. (2002), and a band representing the
EGRET unidentified source 3EG J1714-3857, as taken from Reimer \& Pohl
(2002).  Low model densities, \teq{n_p=0.02}cm$^{-3}$, render the
bremsstrahlung and pion decay contributions insignificant at X-ray and
TeV energies, respectively.
}
 \label{fig:spectra_highB}
\end{figure}

Ellison, Slane \& Gaensler (2001) used the hydrodynamics of a wind-shell
interaction to define shock parameters, using these as input into the
non-linear acceleration formalism of Ellison, Berezhko \& Baring (2000)
that was then used to generate photon emission models.  Here, the
hydrodynamic considerations were not explicitly made, but parameters
similar to those of Ellison, Slane \& Gaensler (2001; specifically a
hybrid of their models A, B and C) were adopted to illustrate the main
features.    Spectra from the first sample model are illustrated in
Fig.~\ref{fig:spectra_highB}, together with relevant, published data. This is a 
model with a modest upstream field of \teq{B_u=2\mu}G and compression ratio
\teq{r_{tot}=5}. The field compression of course depends on the shock
obliquity, so here we essentially model an almost perpendicular shock,
so that \teq{B_d=10\mu}G.

% Figure 1 goes here

The upstream particle density is \teq{n_p=0.02}cm$^{-3}$ for this
example, low enough that a pion decay signal would be dominated by the
inverse Compton one in the TeV band (see Fig.~3 of Ellison, Slane \&
Gaensler 2001). For the same reason, the bremsstrahlung contribution
would lie far below that of synchrotron emission in the X-ray band, and
so is also not exhibited. A source distance of 6 kpc and a remnant age of
\teq{\tSNR =10^4}yrs were adopted, following the choice of Ellison,
Slane \& Gaensler (2001) that were based on the contentions of Slane et
al. (1999).  Debate over the actual distance to SNR G347.3-0.5 will be
addressed below.

The spectra in Fig.~\ref{fig:spectra_highB} differ slightly from those
exhibited in Fig.~3 of Ellison, Slane \& Gaensler (2001), most notably
in being a little sharper in the severity of the X-ray and TeV
turnovers, and in the prominence of the spectral breaks in the infra-red
and the 100 MeV band.  Here, the turnovers are structured, broken
power-laws, as opposed to the exponential ones opted for in Ellison,
Slane \& Gaensler (2001). Also, the spectral breaks image the structure
of the underlying electron distributions from the simplified, analytic,
non-linear acceleration model.  Apart from these subtleties, the results
are similar, and it is clear that the model predicts a TeV turnover too
high to accommodate the steep CANGAROO spectrum.

The global structure of the X-ray and TeV gamma-ray spectral peaks can
be quickly understood to be dependent simply on the shock speed, and the
strength of the magnetic field.  For the parameter regimes chosen to
model SNR G347.3-0.5, sufficiently strong \teq{B} is invoked for the
acceleration of electrons (not ions) to be limited by synchrotron
cooling.  In such cases, the maximum energy of electrons for remnants in
the Sedov phase is (for \teq{B > 3\mu}G, e.g., see Baring et al. 1999):
\begin{equation}
E_{\rm max,e}\;\sim\; \dover{60}{\eta} \biggl( \dover{B_d}{3 \mu G} \biggr)^{-1/2}\,
   \biggl( \dover{n_p}{0.01 \hbox{cm}^{-3}} \biggr)^{-1/5}\,
   \biggl( \dover{\tSNR}{10^4 \hbox{yr} } \biggr)^{-1/2}\, \hbox{TeV}\;\; .
 \label{eq:Emax_highB}
\end{equation}
Here \teq{\eta =\lambda /r_g} is the ratio of the diffusive mean free
path to a particle's gyroradius, a parameter that couples to diffusive
transport across fields and that determines the acceleration rate of
electrons and ions. This form guarantees that the X-ray synchrotron turnover
energy (i.e., just above the X-ray spectral peak in the \teq{\nu F_{\nu}} 
representation), which is proportional to \teq{E_{\rm max,e}^2 B_d}, is
independent of the field strength, and depends only on the shock speed
(e.g. see de Jager et al. 1996, and the discussion in de Jager \& Baring
1997) through its dependence on \teq{n_p} and \teq{\tSNR}.

In contrast, the inverse Compton (IC) peak occurs at an energy
\teq{\propto E_{\rm max,e}^2} that does scale with the field \teq{B_d}:
higher fields are needed to move the turnover to sub-TeV energies.  At
the same time, the normalization of the X-ray and TeV peaks is
controlled purely by the relative energy densities in the soft CMB
photons and the downstream magnetic field, an elementary characteristic
(e.g. see Rybicki \& Lightman 1979), since the same electrons are
emitting in each band.   Higher fields enhance the synchrotron component
relative to the IC one, i.e. suppress the gamma-rays if the volume
filling factor \teq{f} (here it is around \teq{f=0.3}) is chosen to fix
the radio to X-ray continuum. Hence, it is quickly established that, in
the \teq{\nu F_{\nu}} representation of Fig.~\ref{fig:spectra_highB}, where the y-axis scales
with the emitted power, the X-ray and TeV gamma-ray peaks will be of the
same height when \teq{B_d =3.3\mu}G. Adjustments in \teq{B_d}, \teq{f}
and \teq{n_p} would be needed to try to match the CANGAROO data in
Fig.~\ref{fig:spectra_highB}.  The very recent HESS detection provides additional challenges,
as discussed below.

\subsection{A Model for Nearby Distances}
 \label{sec:nearby}
 
A crucial ingredient in the model presented in Fig.~\ref{fig:spectra_highB} is the assumption
that the source is 6 kpc distant.   Koyama et al. (1997) obtained a
distance estimate of \teq{D=1}kpc based on the column density inferred
from soft X-ray absorption in ASCA data, when compared to the Galactic
Center direction absorption characteristics.  This contention has
recently been supported by XMM observations (Cassam-Chena\"{i} et al.
2004) and also evidence for molecular cloud material at 1 kpc (Fukui et
al. 2003). Clearly, for the observed expansion rate, the age of the
remnant is therefore uncertain by a factor of 6 (though it would still
be inferred to be in the Sedov phase), and so also is the inferred mean
density in the remnant, because of the relationship between density
\teq{n_p} and distance \teq{D} for fixed source synchrotron flux:
\teq{F\propto n_p D=} constant. Slane et al. (1999) favored the larger 6
kpc value, principally because they observed that CO intensity mapping
suggested a projected association of the remnant with three molecular
clouds that appeared to have CO and HI column densities matching the
X-ray values. The radial velocities of these clouds indicates a minimum
distance of \teq{D=6}kpc when folded into a galactic rotation model. 
This reasoning obviously depends on the validity of the association.

The debate in the literature over this distance has not been
conclusively resolved.   Accordingly, presenting a model for a
\teq{D=1}kpc case is clearly motivated.  This translates to assuming a
density around \teq{n_p\sim 0.2}cm$^{-3}$, for which we will not treat
bremsstrahlung or line emission here.  From Eq.~(\ref{eq:Emax_highB}),
this change in density would force \teq{E_{\rm max,e}} lower so as to
move the synchrotron peak below the X-ray band.   Adjusting \teq{B_d}
upwards is to no avail, since the synchrotron peak energy is independent
of field strength in this cooling-dominated domain. Hence to fit the
ASCA continuum, lower fields are desirable to emerge from
cooling-limited acceleration phase space, and thereby render \teq{E_{\rm
max,e}} sensitive to the choice of \teq{B_d}. For fields \teq{B \lesssim
3\mu}G, synchrotron cooling is generally insignificant, and the maximum
energy of accelerated electrons is controlled by the spatial scale of
the remnant (e.g. Baring et al. 1999): 
\begin{equation}
E_{\rm max,e}\;\sim\; \dover{170}{\eta} \biggl( \dover{B_u}{3 \mu G} \biggr)\,
   \biggl( \dover{n_p}{\hbox{cm}^{-3}} \biggr)^{-1/5}\,
   \biggl( \dover{\tSNR}{10^4 \hbox{yr} } \biggr)^{-1/2}\, \hbox{TeV}\;\; .
 \label{eq:Emax_lowB}
\end{equation}
For the higher density and younger age dictated by choosing
\teq{D=1}kpc, a field of \teq{B_d=2\mu}G yields a synchrotron \teq{\nu
F_{\nu}} peak in the classical X-ray band.

\begin{figure}[htb]
\figureoutpdf{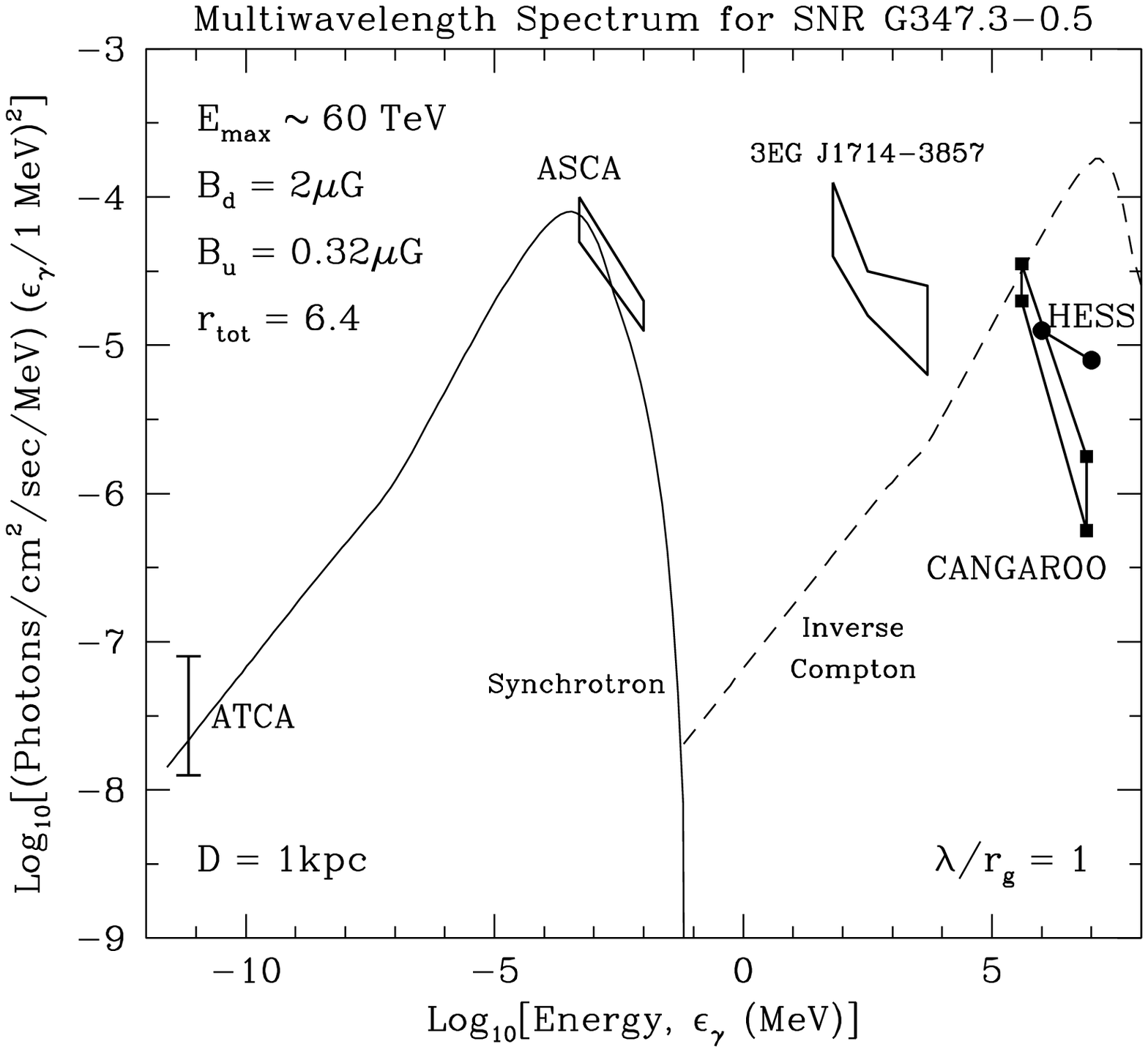}{5.5}{0.0}{-0.25}{
The multi-wavelength model spectrum resulting from a non-linear shock
acceleration model of SNR G347.3-0.5 (RX J1713.7-3946) for an assumed
distance of \teq{D=1}kpc, following the proposition of Koyama et al.
(1997).  The two spectral components are synchrotron emission (solid
curve) and inverse Compton scattering (dashed).  Model and source
parameters are as labeled.   The observational data are as in Fig.~\ref{fig:spectra_highB},
with the addition of the \teq{E^{-2.2}} spectrum for the TeV detection
recently announced by the HESS telescope team (Aharonian et al. 2004);
see text for a discussion on normalization.
}
 \label{fig:spectra_lowB}
\end{figure}

A model spectrum corresponding to this case is depicted in Fig.~\ref{fig:spectra_lowB}.  The
compression ratio of \teq{r_{tot}=6.4} was slightly higher than in the
first example (due in part to a higher electron-to-proton abundance
ratio and a higher value of \teq{E_{\rm max,e}} that spawns increased
losses in the non-linear acceleration model), and again represented
\teq{B_d/B_u} as in a quasi-perpendicular shock.  The maximum electron
(and proton) energy was 60 TeV in this case, markedly higher than in
Fig.~\ref{fig:spectra_highB}.  Consequently, the inverse Compton \teq{\nu F_{\nu}} peak moves
to above 10 TeV, a gross mismatch to the putative CANGAROO points.  The
flux in the IC peak now exceeds that for the X-ray synchrotron peak,
since now the magnetic field energy density is inferior to that of the
cosmic microwave background.

% Figure 2 goes here

Also marked on the Figure is the spectrum announced by the HESS team in
Aharonian et al. (2004).  This spectrum is very different from the
CANGAROO one.  Not only is it flatter (\teq{dn/dE\propto E^{-2.2}} as
opposed to \teq{dn/dE\propto E^{-2.8}}), it possesses slightly higher
overall flux.  The HESS result is integrated over the entire remnant,
and therefore differs in normalization from the CANGAROO result, which
was for the northwest rim portion only. For the purposes of the model
comparison, the HESS spectrum from just this confined portion of the rim
is somewhat lower than the CANGAROO data, reflecting the sensitive
nature of the HESS telescope array. It is noteworthy that the HESS
detection is spatially resolved, mirroring the ASCA sky map more closely
due to its improved angular resolution.  Clearly the HESS results are at
odds with the CANGAROO data, a topical issue that will be the subject of
discussion by the TeV astronomy community in the coming years. 
Moreover, the HESS data are just as difficult to reconcile with the
model spectrum, as depicted.

This multiwavelength modeling could be viewed as another argument
against adopting \teq{D=1}kpc for the distance to SNR G347.3-0.5.
However, it should be noted that the models were restricted to the Bohm
diffusion limit \teq{\eta =\lambda /r_g=1}, where a particle's diffusive
scale is the smallest possible, i.e. its gyroradius, corresponding to
strong field turbulence.  Therefore, less turbulent shock environments
can increase the diffusive scale, thereby reducing the acceleration rate
and lowering \teq{E_{max,e}} (see Eqs.~(\ref{eq:Emax_highB})
and~(\ref{eq:Emax_lowB}) ) without necessarily altering other
environmental parameters such as \teq{B_d} and \teq{n_p}.  Modifications
along these lines can, with a slight adjustment upwards in field
strength, approximately match the CANGAROO data in Fig.~\ref{fig:spectra_lowB}, but might
have greater difficulty in accommodating the portion of the HESS signal
coming from the northwest rim, which would require higher fields in
order to reduce the inverse Compton to synchrotron ratio at their
respective peaks.

\subsection{Hadronic Models?}
 \label{sec:hadronic}

Having explored the case for purely leptonic models, it is worth noting
that hadronic models for the generation of gamma-rays are in principal
tenable for SNR G347.3-0.5, though not in the simplest form envisaged by
Enomoto et al. (2002) in their original claim.  The objections of Reimer
\& Pohl (2002) and Butt et al. (2002) still stand.  However the complex
environment of the remnant admits the possibility that pions can be
created using molecular cloud nucleons as targets for the energetic
cosmic rays, and these would be somewhat remote from the expanding outer
shell of SNR G347.3-0.5.  Such a departure from a one-zone scenario is
quite reasonable in these circumstances.

A spatial separation of accelerator and target alters the expected
spectrum for pion decay emission.  Since the diffusive scale of the
higher energy ions is larger than that for the low energy ones that
contribute to the \teq{\pi^0} decay peak at 67 MeV (e.g. see Jones \&
Ellison 1991; Baring et al. 1999; Berezhko \& Ellison 1999, for a
discussion of such diffusion lengths), the latter ions have great
difficulty in convecting far upstream of the shocked shell.
Consequently, the pion signal would be strongly suppressed near 67 MeV
in this scenario, as would all such emission from particles whose
energies correspond to diffusive scales less than the distance between
the shell and the target cloud. This would skew the \teq{\pi^0} decay
spectrum to render it peaked near the TeV band in the \teq{\nu F_{\nu}}
representation, thereby evading the EGRET source bounds.  

Clearly, for this geometrical separation to have a significant impact on
the pion decay continuum, the distance between remnant shell and target
molecular cloud must be at least a sizable fraction of the radius of the
remnant. Such a model is not unduly contrived, but does require
additional parameters relating to spatial scales and turbulence (i.e.
\teq{\eta =\lambda /r_g}) that are poorly constrained by observational
data, particularly given the impact of line-of-sight ambiguities to the
environmental geometry.  Furthermore, these parameters might well have
to vary significantly across the remnant in order to reconcile the
molecular cloud sky geometries with the X-ray and TeV emission
morphologies.  Finally, it might prove difficult to distinguish this
pion decay scenario spectroscopically from inverse Compton emission,
given that both would be extremely flat and statistics-limited somewhat
below the TeV peak.

\section{Conclusions}
 \label{sec:conclusion}

In closing, two multi-wavelength models are presented here for broadband
emission in SNR G347.3-0.5, using fundamental characteristics of
non-linear shock acceleration as their basis.  This offering is not
designed to find the ``perfect'' model fit, but rather to represent
samples from an array of possibilities that indicate how such
multi-wavelength approaches can constrain model parameter space.  While
the examples are focused on an inverse Compton origin of TeV emission,
which is the preferred interpretation in a low density remnant
environment, it is also emphasized that hadronic models with pion decay
emission in the TeV band are not presently ruled out.  Using nucleonic
targets in proximate molecular clouds that are distinct from the remnant
shell can potentially generate acceptable model fits to TeV datasets.

The study of this remnant is presently burdened with the controversy
concerning the TeV data, an issue also for SN 1006, which has so far not
been detected by HESS (Aharonian et al. 2005). The contradictory results
for SNR G347.3-0.5 between CANGAROO and HESS obviously need to be
resolved in order to propel theoretical understanding at a more rapid
pace.  This is a challenge that the TeV community met in the case of the
Crab nebula, and northern hemisphere blazars, that established the
branch of TeV astronomy on a firm basis. For supernova remnants, this
bridge has yet to be crossed. Anticipating watershed results from the
increasing number of IACTs to be in operation in both hemispheres over
the next few years, such a development will hopefully be around the
corner.

\section{References}

% \reference[Names(Year)]{label} or \reference[Names(Year)Long names]{label}.
% (\harvarditem{Name}{Year}{label} is also supported.)

\setlength{\parskip}{.00in}

\reference
Aharonian, F., Akhperjanian,ÊA.~G., Barrio,ÊJ., et al. Evidence for TeV gamma ray 
   emission from Cassiopeia A. \aa\vol{370}{112--120} 2001.
\reference
Aharonian, F., Akhperjanian,ÊA.~G., Aye,ÊK.-M., et al. High-energy particle 
   acceleration in the shell of a supernova remnant. \nat\vol{432}{75--77} 2004.
\reference
Aharonian, F., Akhperjanian,ÊA.~G., Aye,ÊK.-M., et al. Upper limits to the SN1006
   multi-TeV gamma-ray flux from H.E.S.S. observations. \aa\ in press. 2005.
\reference
Baring, M.~G., Ellison,ÊD.~C., Reynolds,ÊS.~P., et al. Radio to Gamma-Ray 
   Emission from Shell-Type Supernova Remnants: Predictions from Nonlinear 
   Shock Acceleration Models. \apj\vol{513}{311--338} 1999.
\reference
Baring, M.~G. Modelling Hard Gamma-Ray Emission From Supernova Remnants.
   in \it GeV--TeV Gamma-Ray Astrophysics Workshop,
   \rm ed. B.~L. Dingus, M.~H. Salamon, \& D.~B. Kieda
   (AIP Conf. Proc. 515, New York), pp.~173--182. {\tt [astro-ph/9911060]} 2000.
\reference
Berezhko, E.~G., Ellison, D.~C. A Simple Model of Nonlinear Diffusive 
   Shock Acceleration. \apj\vol{526}{385--399} 1999.
\reference
Berezhko, E.~G., Yelshin, V., Ksenofontov, L.~T. Cosmic Ray Acceleration
   in Supernova Remnants. \jetp\vol{82}{1--21} 1996.
\reference
Blasi, P. A semi-analytical approach to non-linear shock acceleration.
   \app\vol{16}{429--439} 2002.
\reference
Butt, Y.~M., Torres,ÊD.~F., Romero,ÊG.~E., et al. Supernova-Remnant Origin 
   of Cosmic Rays? \nat\vol{418}{499--499} 2002.
\reference
Cassam-Chena\"{i}, G., Decourchelle, A., Ballet, J., et al. XMM-Newton observations 
   of the supernova remnant RX J1713.7-3946 and its central source observations 
   of SNR RX J1713.7-3946. \aa\vol{427}{199--216} 2004.
\reference
Decourchelle, A., Ellison, D.~C., Ballet, J. Thermal X-Ray Emission and Cosmic-Ray 
   Production in Young Supernova Remnants. \apj\vol{543}{L57--L60} 2000.
\reference
de Jager, O.~C., Baring, M.~G. Supernova Remnants and Plerions in the 
   Compton Gamma-Ray Observatory Era. in {\it Proc. 4th Compton Symposium,} 
   ed. Dermer, C.~D. \& Kurfess, J.~D. (AIP Conf. Proc. 410, New York), pp.~171--180.
   {\tt [astro-ph/9711212]} 1997.
\reference
de Jager, O.~C., Harding,ÊA.~K., Michelson,ÊP.~F., et al. Gamma-Ray Observations 
   of the Crab Nebula: A Study of the Synchro-Compton Spectrum. 
   \apj\vol{457}{253--266} 1996.
\reference
Drury, L.~O'C. An Introduction to the Theory of Diffusive Shock Acceleration of 
   Energetic Particles in Tenuous Plasmas.  Rep. Prog. Phys. \vol{46}{973--1028} 1983.
\reference
Drury, L.~O'C., Aharonian, F.~A., V\"olk, H.~J. The gamma-ray visibility of 
   supernova remnants. A test of cosmic ray origin. \aap\vol{287}{959--971} 1994.
\reference
Ellison, D.~C., Baring, M.~G., Jones, F.~C. Nonlinear Particle Acceleration 
   in Oblique Shocks. \apj\vol{473}{1029--1050} 1996.
\reference
Ellison, D.~C., Berezhko, E.~G., Baring, M.~G. Nonlinear Shock Acceleration 
   and Photon Emission in Supernova Remnants. \apj\vol{540}{292--307} 2000.
\reference
Ellison, D.~C., Decourchelle, A., Ballet, J. Nonlinear particle acceleration 
   at reverse shocks in supernova remnants. \aap\vol{429}{569--580} 2005.
\reference
Ellison, D.~C., Eichler, D. Monte Carlo shock-like solutions to the Boltzmann 
   equation with collective scattering. \apj\vol{286}{691--701} 1984.
\reference
Ellison, D.~C., Slane, P.~O., Gaensler, B.~M. Broadband Observations 
   and Modeling of the Shell-Type Supernova Remnant G347.3-0.5. 
   \apj\vol{563}{191--201} 2001.
\reference
Enomoto, R., Tanimori,ÊT., Naito,ÊT., et al. The acceleration of cosmic-ray protons 
   in the supernova remnant RX J1713.7-3946. \nat\vol{416}{823--826} 2002.
\reference
Fukui,ÊY., Moriguchi,ÊY., Tamura,ÊK., et al. Discovery of Interacting Molecular Gas
   toward the TeV Gamma-Ray Peak of the SNR G 347.3-0.5. \pasj\vol{55}{L61--L64} 2003.
\reference
Gaisser, T.~K., Protheroe, R.~J., Stanev, T.  Gamma-Ray Production in 
   Supernova Remnants\apj\vol{492}{219--227} 1998.
\reference
Jones, F.~C., Ellison, D.~C. The plasma physics of shock acceleration. 
   \ssr\vol{58}{259--346} 1991.
\reference
Kang, H., Jones, T.~W. Diffusive Shock Acceleration Simulations: Comparison 
   with Particle Methods and Bow Shock Measurements. \apj\vol{447}{944--961} 1995.
\reference
Koyama, K., Petre,ÊR., Gotthelf,ÊE.~V., et al. Evidence for Shock Acceleration of 
   High-Energy Electrons in the Supernova Remnant SN:1006. \nat\vol{378}{255--257} 1995.
\reference
Koyama, K., Kinugasa,ÊK., Matsuzaki,ÊK., et al. Discovery of Non-Thermal X-Rays 
   from the Northwest Shell of the New SNR RX J1713.7-3946: The Second 
   SN 1006? \pasj\vol{49}{L7--L11} 1997.
\reference
Lazendic,ÊJ.~S., Slane,ÊP.~O., Gaensler,ÊB.~M., et al. A High-Resolution Study of 
   Nonthermal Radio and X-Ray Emission from Supernova Remnant G347.3-0.5.
   \apj\vol{602}{271--285} 2004.
\reference
Malkov, M.~A. Analytic Solution for Nonlinear Shock Acceleration in the 
   Bohm Limit. \apj\vol{485}{638--654} 1997.
\reference
Reimer, O., Pohl, M. No evidence yet for hadronic TeV gamma-ray emission 
   from SNR RX J1713.7-3946. \aa\vol{390}{L43--L46} 2002.
\reference
Rybicki, G.~B., Lightman, A.~P.  \rm Radiative Processes in
   Astrophysics \rm (Wiley, New York) 1979.
\reference
Slane, P.~O., Gaensler,ÊB.~M., Dame,ÊT.~M., et al. Nonthermal X-Ray Emission 
   from the Shell-Type Supernova Remnant G347.3-0.5\apj\vol{525}{357--367} 1999.
\reference
Sturner, S.~J., Skibo, J.~G., Dermer, C.~D., Mattox, J.~R. Temporal Evolution of 
   Nonthermal Spectra from Supernova Remnants. \apj\vol{490}{619--632} 1997.
\reference
Tanimori, T., Hayami,ÊY., Kamei,ÊS., et al. Discovery of TeV Gamma Rays 
   from SN 1006: Further Evidence for the Supernova Remnant Origin 
   of Cosmic Rays. \apj\vol{497}{L25--L28} 1998.

\end{document}